\pdfoutput=1

\documentclass[11pt,twoside,a4paper,cmspaper,final,collab]{cms-tdr}
\def\svnVersion{7ae6751-D}\def\svnDate{2020/10/09}\def\cmsCernNoTag{CERN-EP-2020-070}\def\cmsCernDate{\today}\def\cmsMessage{"Published in Physical Review Letters as \href{http://dx.doi.org/10.1103/PhysRevLett.125.152001}{\doi{10.1103/PhysRevLett.125.152001}.}"}

\begin{document}\cmsNoteHeader{BPH-17-005}

\hyphenation{had-ron-i-za-tion}
\hyphenation{cal-or-i-me-ter}
\hyphenation{de-vices}
\providecommand{\Pxh}{{\HepParticle{h}{}{}}\Xspace} 
\providecommand{\PX}{{\HepParticle{X}{}{}}\Xspace} 
\newcommand{\mumu}      {\ensuremath{\PGmp\PGmm}\xspace}
\newcommand{\pip}       {\PGpp}
\newcommand{\pim}       {\PGpm}
\newcommand{\pipm}      {\PGppm}

\newcommand{\kaon}      {\ensuremath{\PK}\xspace}
\newcommand{\Kbar}      {\PAK}

\newcommand{\Kp}        {\PKp}
\newcommand{\Km}        {\PKm}
\newcommand{\Kz}        {\PKz}
\newcommand{\Kzb}       {\PAKz}
\newcommand{\Kpm}       {\PKpm}

\newcommand{\Bu}        {\PBp}
\newcommand{\Bpm}       {\PBpm}
\newcommand{\Bd}        {\PBz}
\newcommand{\Bs}        {\PBzs}

\newcommand{\jpsi}      {\PJGy}
\newcommand{\psit}      {\Pgy}
\newcommand{\Xm}        {\HepParticleResonance{X}{3872}{}{}\Xspace}

\newcommand{\pipi}      {\ensuremath{\pip\pim}\xspace}
\newcommand{\JPP}       {\ensuremath{\PJGy\hspace{.06em}\pipi}\xspace}
\newcommand{\PJPP}      {\ensuremath{\psit\to\JPP}\xspace}
\newcommand{\XmJPP}     {\ensuremath{\Xm\to\JPP}\xspace}
\newcommand{\mJPP}      {\ensuremath{m(\JPP)}\xspace}
\newcommand{\PKK}       {\ensuremath{\PGf\to\Kp\Km}\xspace}
\newcommand{\KK}        {\ensuremath{\Kp\Km}\xspace}
\newcommand{\mKK}       {\ensuremath{m(\KK)}\xspace}

\newcommand{\JKKPP}     {\ensuremath{\PJGy\hspace{.06em}\KK\pipi}\xspace}
\newcommand{\mJKKPP}    {\ensuremath{m(\JKKPP)}\xspace}
\newcommand{\BsPP}      {\ensuremath{\Bs\to\psit\PGf}\xspace}
\newcommand{\BsXP}      {\ensuremath{\Bs\to\Xm\PGf}\xspace}

\newcommand{\pp}        {\ensuremath{\Pp\Pp}\xspace}
\newcommand{\mjpsipdg}  {\ensuremath{m^{\text{PDG}}_{\PJGy}}\xspace}
\newcommand{\splot}     {\ensuremath{{}_{s}\mathcal{P}\mathrm{lot}}\xspace}
\newcommand{\ebran}     {\ensuremath{\,(\mathcal{B})}\xspace}

\ifthenelse{\boolean{cms@external}}{\providecommand{\cmsLeft}{upper\xspace}}{\providecommand{\cmsLeft}{left\xspace}}
\ifthenelse{\boolean{cms@external}}{\providecommand{\cmsRight}{lower\xspace}}{\providecommand{\cmsRight}{right\xspace}}
\newlength\cmsTabSkip\setlength{\cmsTabSkip}{1ex}

\ifthenelse{\boolean{cms@external}}
{\providecommand{\suppMaterial}{the Supplemental Material~\cite{suppMatBib}}}
{\providecommand{\suppMaterial}{Appendix~\ref{app:suppMat}\nocite{suppMatBib}}}

\cmsNoteHeader{BPH-17-005} 
\title{Observation of the \texorpdfstring{$\BsXP$}{B0s to X(3872) phi} decay}
\date{\today}

\abstract{
Using a data sample 
of proton-proton collisions at $\sqrt{s}= 13\TeV$,
corresponding to an integrated luminosity of 140\fbinv
collected by the CMS experiment in 2016--2018, 
the \BsXP decay is observed.
Decays into \JPP and \KK are used to reconstruct, respectively, 
the \Xm and $\PGf$.
The ratio of the product of branching fractions
$\mathcal{B}(\BsXP)\,\mathcal{B}(\XmJPP)$
to the product $\mathcal{B}(\BsPP)\,\mathcal{B}(\PJPP)$
is measured to be $(2.21\pm0.29\stat\pm0.17\syst)\%$.
The ratio $\mathcal{B}(\BsXP)/\mathcal{B}(\Bd\to\Xm\Kz)$ is found to be consistent 
with one, while the ratio $\mathcal{B}(\BsXP)/\mathcal{B}(\Bu\to\Xm\Kp)$ is 
two times smaller. This suggests a difference in the production dynamics
of the \Xm in \Bd and \Bs meson decays compared to \Bu. 
The reported observation may shed new 
light on the nature of the \Xm particle.
}

\hypersetup{%
pdfauthor={CMS Collaboration},%
pdftitle={Observation of the B0s to X(3872) phi decay},%
pdfsubject={CMS},%
pdfkeywords={CMS, bottom quark,  X(3872), exotic state, exotic meson, charmonium}}

\maketitle 

The observed spectrum of $\PQc\PAQc$ states below the $\PD\PAD$ threshold agrees well with 
theoretical predictions~\cite{Voloshin:2007dx,Brambilla:2010cs}.
Since the advent of the BaBar and Belle experiments at the B factories and their 
discovery of several charmonium-like states, the conventional charmonium 
model above the $\PD\PAD$ threshold has become the subject of intense discussions.
In 2003, the Belle Collaboration 
observed a new particle in the $\Bu\to\JPP\Kp$ decay~\cite{belle}, named \Xm and decaying
to \JPP, with a very small natural width 
for a state above the $\PD\PAD$ threshold. 
Its world-average mass is $3871.69\pm0.17\MeV$, which is extremely 
close to the $\PADz\PDst^0$ threshold of $3872.68\pm0.07\MeV$~\cite{PDG}. 
With this mass and a total width less than $2\MeV$~\cite{allaboutrho,Aaij:2020qga},
the \Xm particle did not match any 
of the theoretically predicted charmonium resonances.

The discovery of \Xm opened a new era of exotic, quarkonium-like spectroscopy. 
Many new states with unusual properties have been observed, including several 
charged states~\cite{Brambilla:2019esw,PDG}.
At hadron colliders, prompt processes were found to be the dominant \Xm
production mechanism~\cite{Bauer:2004bc,Chatrchyan:2013cld,Aaboud:2016vzw}. 
The nature of \Xm, also known as \PGccDoP{3872},
is still unexplained in spite of the determination of its quantum numbers 
($J^{PC}=1^{++}$)~\cite{Abulencia:2006ma,lhcb,Aaij:2015eva}. The studies of the dipion mass 
spectrum~\cite{Abulencia:2005zc,Abulencia:2006ma,allaboutrho,Chatrchyan:2013cld,lhcb,Aaij:2015eva,Aaboud:2016vzw}
clearly favor the presence of the intermediate \PGrzP{770} state in the 
isospin violating $\XmJPP$ decay. 
Important information about the \Xm production in weak 
decays can be extracted by comparing the branching fractions 
$\mathcal{B}(\PB\to\Xm\Pxh)$
for different \PB mesons, where $\Pxh$ denotes a light hadron. 
More measurements of \PQb hadron 
decays involving \Xm production would provide important inputs 
for understanding its internal structure and creation dynamics.

This Letter reports the first observation of the \BsXP decay, 
where \XmJPP and \PKK decays are used to reconstruct the intermediate resonances,
and the measurement of the following ratio of branching fractions:
\ifthenelse{\boolean{cms@external}}
{
\begin{linenomath}
\begin{equation} 
\begin{aligned}
R & \equiv \frac{\mathcal{B}(\BsXP)\,\mathcal{B}(\XmJPP)}{\mathcal{B}(\BsPP)\,\mathcal{B}(\PJPP)}\\
  & = \frac{N(\BsXP)}{N(\BsPP)}\,\frac{\epsilon_{\BsPP}}{\epsilon_{\BsXP}}.\label{eqr}
\end{aligned}
\end{equation} 
\end{linenomath}
}
{
\begin{linenomath}
\begin{equation} 
\begin{split}
R  \equiv \frac{\mathcal{B}(\BsXP)\,\mathcal{B}(\XmJPP)}{\mathcal{B}(\BsPP)\,\mathcal{B}(\PJPP)}
   = \frac{N(\BsXP)}{N(\BsPP)}\,\frac{\epsilon_{\BsPP}}{\epsilon_{\BsXP}}.\label{eqr}
\end{split}
\end{equation} 
\end{linenomath}
}
In this expression, $N$ stands for the measured number 
of signal events in data, and $\epsilon$ stands for the efficiency.
The \PJGy and \PGfP{1020} (referred to as $\PGf$ throughout the Letter) mesons are 
reconstructed in the \mumu and \KK channels, respectively. 
The normalization is done via the \BsPP decay, with a subsequent 
\PJPP decay. 
The similarity of the decay topology of the signal and 
normalization channels results in nearly identical kinematics, leading to the 
cancellation of many systematic uncertainties in the ratio.

The central feature of the CMS apparatus~\cite{Chatrchyan:2008zzk} 
is a superconducting solenoid of 
6\unit{m} internal diameter, providing a magnetic field of 3.8\unit{T}.
Within the solenoid volume are a silicon pixel and strip tracker, a lead 
tungstate crystal electromagnetic calorimeter, and a brass and scintillator 
hadron calorimeter. Muons are detected in gas-ionization chambers embedded in 
the steel flux-return yoke outside the solenoid. 
The analysis uses proton-proton (\pp) collision data recorded by the CMS detector during 
the LHC Run 2 in 2016--2018 at $\sqrt{s}=13\TeV$, corresponding to an 
integrated luminosity of $140\fbinv$. 
Events of interest are selected using a two-tiered trigger system~\cite{Khachatryan:2016bia}. 
The first level (L1), composed of custom hardware processors, uses information from the 
calorimeters and muon detectors to select events at a rate of around 100\unit{kHz} within a time 
interval of less than 4\mus. The L1 trigger used in the analysis requires at least two muons. 
The second level, known as the high-level trigger, consists of a farm of processors running 
a version of the full event reconstruction software optimized for fast processing that reduces the 
event rate to around 1\unit{kHz} before data storage. The high-level trigger algorithm used in the analysis 
requires two opposite-sign (OS) muons compatible with the dimuon decay of a \PJGy meson at 
a significant distance from the beam axis, as well as an additional track with
transverse momentum $\pt>1.2\GeV$, compatible with being produced in the dimuon vertex.

Simulated event samples for the \BsXP and \BsPP decays are generated in the analysis. 
The \PYTHIA~8.230 package~\cite{Pythia8} is used to simulate the production of the \Bs mesons,
which are subsequently decayed with \EVTGEN 1.6.0~\cite{Lange2001152}, where the final-state photon 
radiation is included using \PHOTOS 3.61~\cite{BARBERIO1991115,BARBERIO1994291}.
Generated events are then passed to a detailed \GEANTfour-based
simulation~\cite{Agostinelli2003250} of the CMS detector, followed by the
same trigger and reconstruction algorithms as used for the collision data.
The simulation includes effects from multiple \pp interactions in the same
or nearby bunch crossings (pileup) with the multiplicity distribution tuned to match the data.

The event selection begins by requiring two OS muons with $\pt>4\GeV$ 
passing the soft-muon identification criteria~\cite{Chatrchyan:2012xi} 
and matching those that triggered the event readout.
The dimuon mass is required to be compatible with the world-average 
\PJGy mass~\cite{PDG}, $\mjpsipdg$.
The $\Bs\to\JKKPP$ candidates are obtained by combining the selected \PJGy
candidate with four high-purity tracks~\cite{Chatrchyan:2014fea} with a
total charge of zero that are not matched with the selected muons.
At least one of the four tracks is required to have $\pt>1.2\GeV$ and a
transverse impact parameter significance greater than 2 to match the trigger requirement.
A kinematic vertex fit that constrains the dimuon invariant mass to $\mjpsipdg$ 
is performed on the two muons and four tracks.
From all reconstructed \pp collision vertices, the primary vertex 
is chosen as the one with the smallest pointing angle, as done in 
Refs.~\cite{BPH16002, BPH16003, BPH18005, BPH19002, BPH18007, BPH19003}.
The pointing angle is the angle between the \Bs candidate momentum and 
the vector joining the primary vertex and the reconstructed \Bs candidate decay vertex.
Signal events are eventually selected based on the corrected 
invariant mass $m(\Bs)=\mJKKPP-\mJPP+m^{\text{PDG}}_{\psit/\Xm}$,
where $m^{\text{PDG}}_{\psit}$ and $m^{\text{PDG}}_{\Xm}$ are the world-average 
\psit and \Xm masses, respectively.
This approach ensures the independence between the reconstructed \Bs and 
\JPP masses and improves the \Bs mass resolution.

To select the $\Bs\to\JKKPP$ candidates, 
one must choose which two OS tracks are from kaons, with the other two tracks then 
being associated with pions.
Since the decays of interest have narrow intermediate states \PKK and 
either \psit or \XmJPP, the following criteria are used to assign the tracks 
for the selected $\JKKPP$ candidates: 
\begin{itemize}
\item $3.60<\mJPP<3.95\GeV$
\item $1.00<\mKK<1.04\GeV$ 
\item $5.32<m(\Bs)<5.42\GeV$
\item if more than one of the mass assignments passes the three 
selections above, the candidate is discarded.
\end{itemize}
The selected mass windows are wide enough to allow fits to the mass distributions,
while maintaining a selection efficiency above 99\%.

The selection criteria are optimized 
using the Punzi figure of merit~\cite{Punzi:2003bu}, 
which does not rely on the signal normalization. 
Data sidebands are used to estimate the background, and the \BsXP simulated 
sample is used to measure the signal efficiency.
The resulting selection criteria are as follows:
$\pt(\Bs)>10\GeV$, vertex $\chi^2$ fit probability
$P_{\text{vtx}}(\Bs)>7\%$, 
$\pt(\pipm)>0.7\GeV$,
$\text{min}(\pt(\Kpm))>1.5\GeV$, $\text{max}(\pt(\Kpm))>2.2\GeV$, 
and the decay length of the \Bs candidate in the transverse plane 
$L_{xy}(\Bs) > 15 \sigma_{L_{xy}}(\Bs)$, where $\sigma_{L_{xy}}$ is the uncertainty in $L_{xy}$.
Additionally, the cosine of the angle between the transverse momentum of the 
\Bs candidate and the displacement vector must satisfy $\cos(\vec{\pt},\,\vec{L}_{xy})> 0.999$,
and the invariant mass of the two pions is required to be above 
$0.45\,(0.70)\GeV$ in the $\psit\,(\Xm)$ channel.

{\tolerance=1200
The signal yields of the \BsXP and \BsPP decays are extracted using 
a two-dimensional (2D) maximum likelihood fit to the \mJPP 
and \mKK distributions for \Bs candidates in the range $5.32<m(\Bs)<5.42\GeV$. 
The numbers of $\Xm\PGf$ and $\psit\PGf$ signal events from the fit are assumed to 
come solely from the corresponding \Bs decays. A systematic uncertainty related 
to this assumption is evaluated below.
\par}

\begin{figure}[htb]
\centering
    \includegraphics[width=0.48\textwidth]{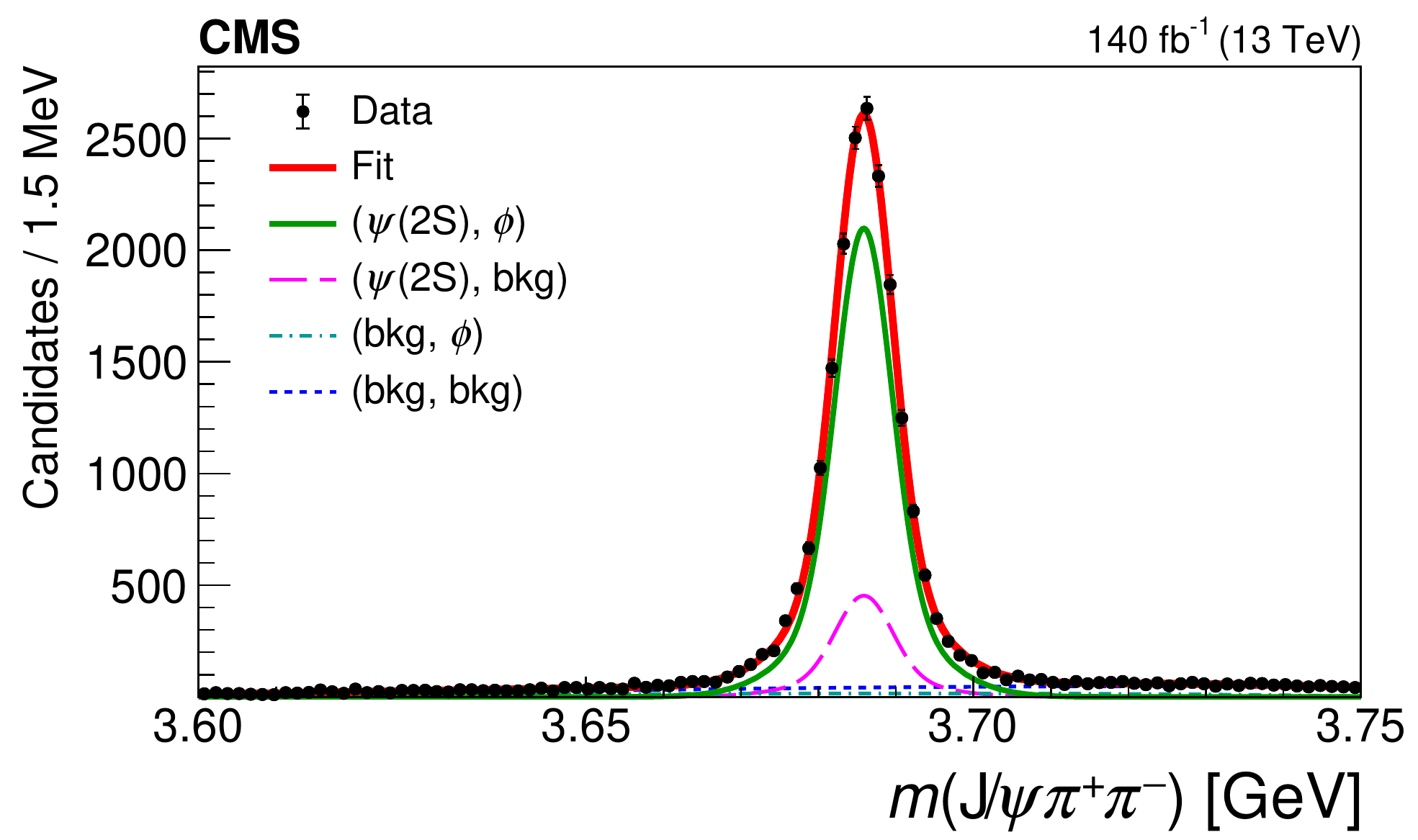}
    \includegraphics[width=0.48\textwidth]{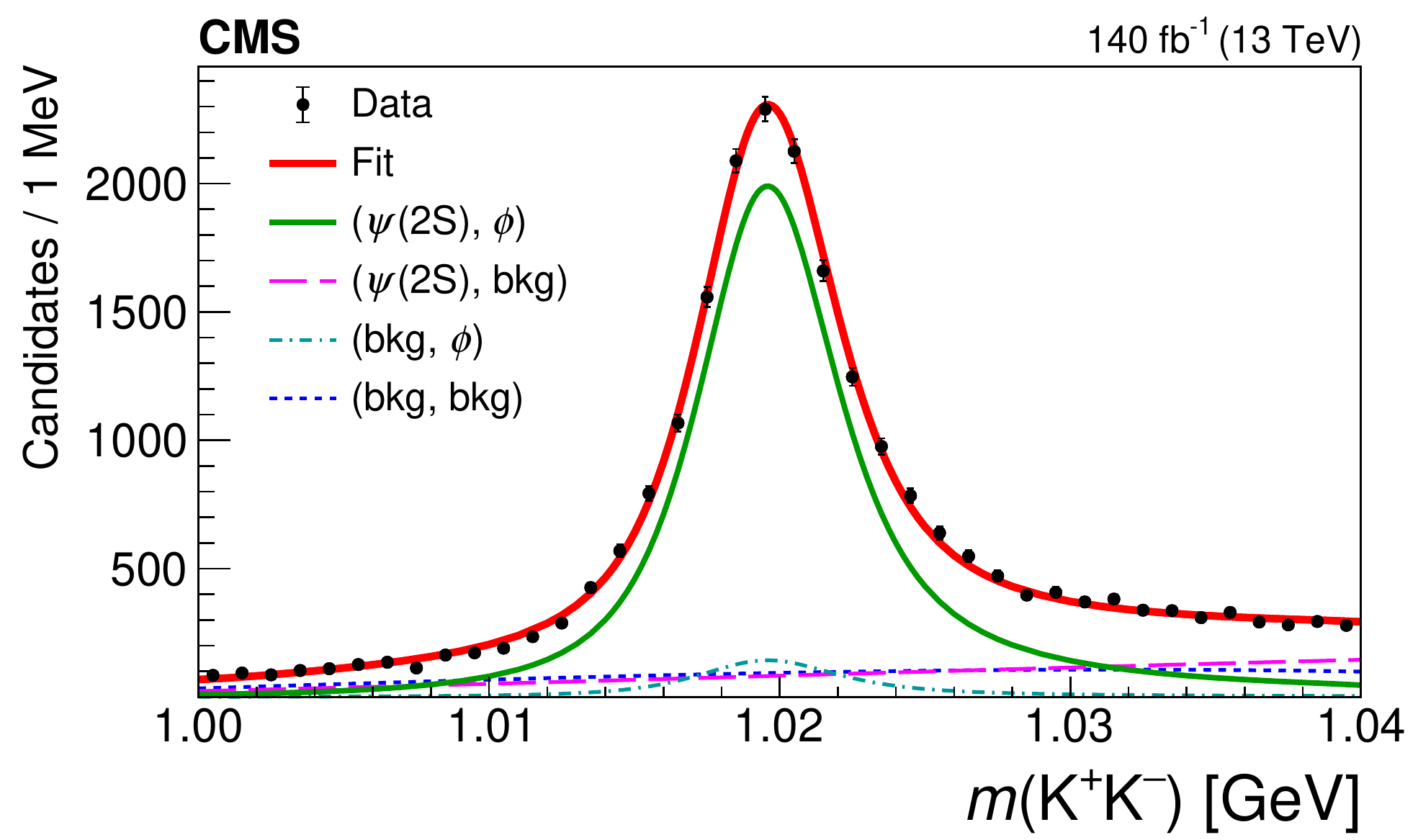}
    \caption{The observed \JPP (\cmsLeft) and \KK (\cmsRight) invariant mass 
    distributions for the \BsPP candidates are shown by the points, with 
    the vertical bars representing the statistical uncertainties. The projections 
    of the 2D fit and its various components are shown by the lines. \label{fig:TwoDfitP}}
\end{figure}

{\tolerance=1200 
Figure~\ref{fig:TwoDfitP} shows the observed \mJPP (\cmsLeft) and 
\mKK (\cmsRight) invariant mass distributions for the 
$\psit\PGf$ candidates with $3.60<\mJPP<3.75\GeV$.
Overlaid are the projections of the 2D fit function, which consists of the 
following four components:
\begin{itemize}
\item $(\psit,\,\PGf)$, for the signal component
\item $(\text{bkg},\,\PGf)$, for events containing genuine \PKK decays and background \JPP combinations
\item $(\psit,\,\text{bkg})$, for events containing genuine \PJPP decays and background \KK combinations
\item $(\text{bkg},\,\text{bkg})$, for the background in both dimensions.
\end{itemize}
Each component is a product of two one-dimensional functions.
For the \PKK signal, a relativistic Breit--Wigner function convolved with the 
detector mass resolution is used, where the $\PGf$ natural 
width is fixed to its known value~\cite{PDG}. 
The mass resolution is determined from simulated event samples to be about $1.3\MeV$.
The background in the \KK mass distribution is modeled with a threshold function multiplied by 
a first-order polynomial: $(\mKK-x_0)^\alpha{Pol_1(\mKK)}$, 
where $x_0$ is the threshold value equal to twice the kaon mass and $\alpha$ is a free parameter. 
The \PJPP signal is described with a double-Gaussian (DG) function with all parameters left free.
The background in the \mJPP distribution is modeled with a modified threshold function: 
$(\mJPP-y_0)^\beta{Pol_1(\mJPP)}$, where $y_0$ is the threshold value equal to 
$\mjpsipdg+0.45\GeV$ (corresponding to the requirement $m(\pipi)>0.45\GeV$), and 
$\beta$ is a free parameter. 
\par}

The following parameters are free in the fit: 
numbers of events in the four components, $\PGf$ and \psit meson masses, 
\psit resolution parameters, 
and background parameters of \mKK and \mJPP.
The fitted yield for the $\psit+\PGf$ component is $N(\BsPP)=15\,359\pm171$. 

\begin{figure}[htb]
\centering
    \includegraphics[width=0.48\textwidth]{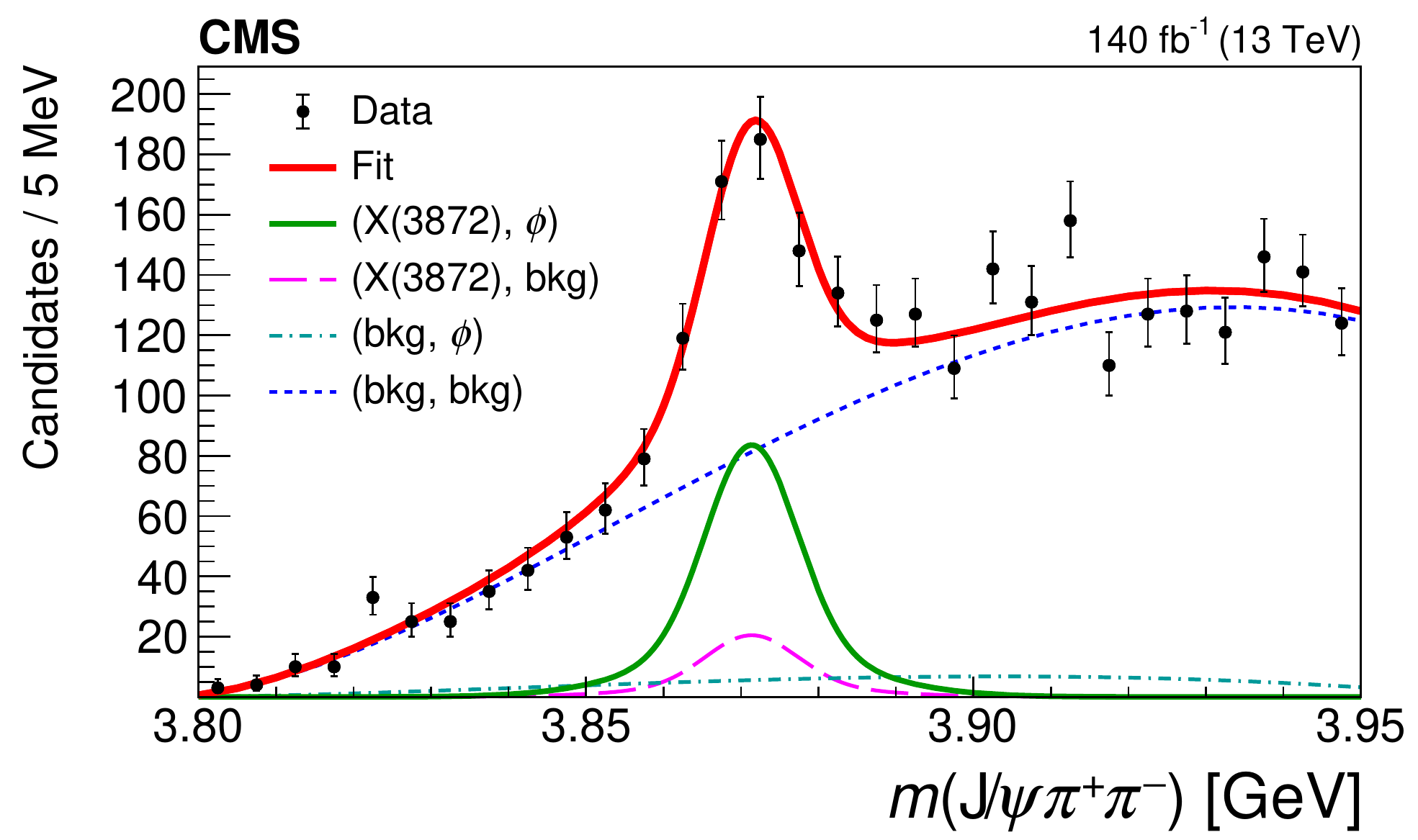}
    \includegraphics[width=0.48\textwidth]{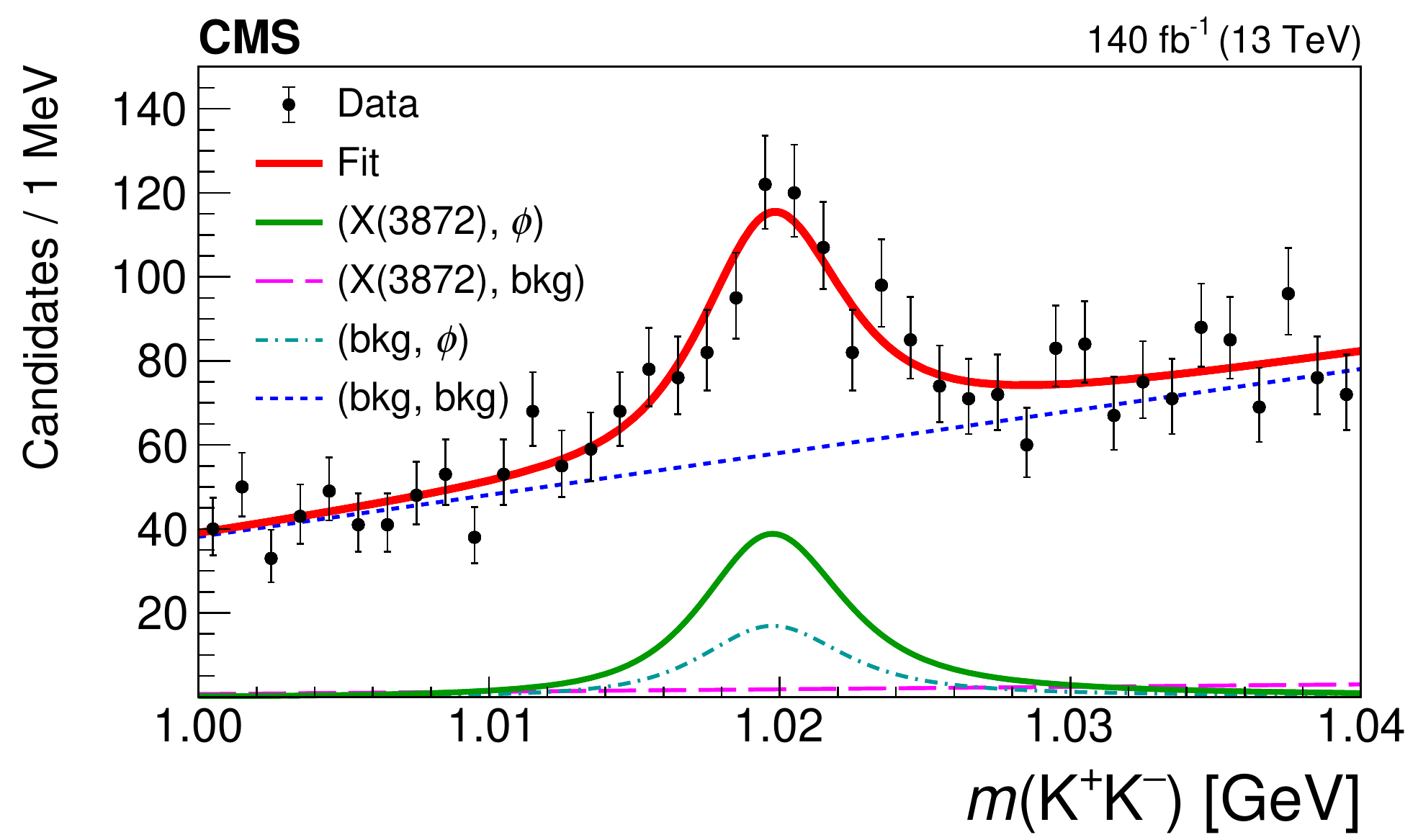}
    \caption{ The observed \JPP (\cmsLeft) and \KK (\cmsRight) invariant mass 
    distributions for the \BsXP candidates are shown by the points, with 
    the vertical bars representing the statistical uncertainties. The projections 
    of the 2D fit and its various components are shown by the lines. \label{fig:TwoDfitX}}
\end{figure}

For the \Xm mass region, defined as $3.80<\mJPP<3.95\GeV$, the same fit 
function is used as in the \psit channel, 
but additional constraints are made because of the lower number of signal events.
The shape of the \XmJPP signal is fixed to the one obtained in data for \PJPP, with 
one floating parameter responsible for the resolution scaling. The \Xm mass
is left free in the fit and the returned value is in agreement with the 
known mass~\cite{PDG}.
The threshold value $y_0$ is changed to $\mjpsipdg+0.7\GeV$ to account for the 
different requirement on the dipion invariant mass applied in the \Xm channel.
The invariant mass distributions and the projections of the 2D fit are shown 
in Fig.~\ref{fig:TwoDfitX}. 
Additional projections of the 2D fit in different 
ranges of \mJPP and \mKK are presented in \suppMaterial.
The measured signal yield is $N(\BsXP)=299\pm39$. 

The statistical significance of the $\BsXP$ signal has been evaluated with the 
likelihood ratio technique by applying the background-only and signal-plus-background 
hypotheses. Using the standard asymptotic 
approximation~\cite{Cowan:2010js} for the likelihood, since the conditions of the 
Wilks' theorem~\cite{Wilks:1938dza} are satisfied,
the statistical significance of the \BsXP signal is over 6 standard deviations ($\sigma$)
after accounting for the systematic uncertainties discussed later.

\begin{figure}[htb]
\centering
    \includegraphics[width=0.48\textwidth]{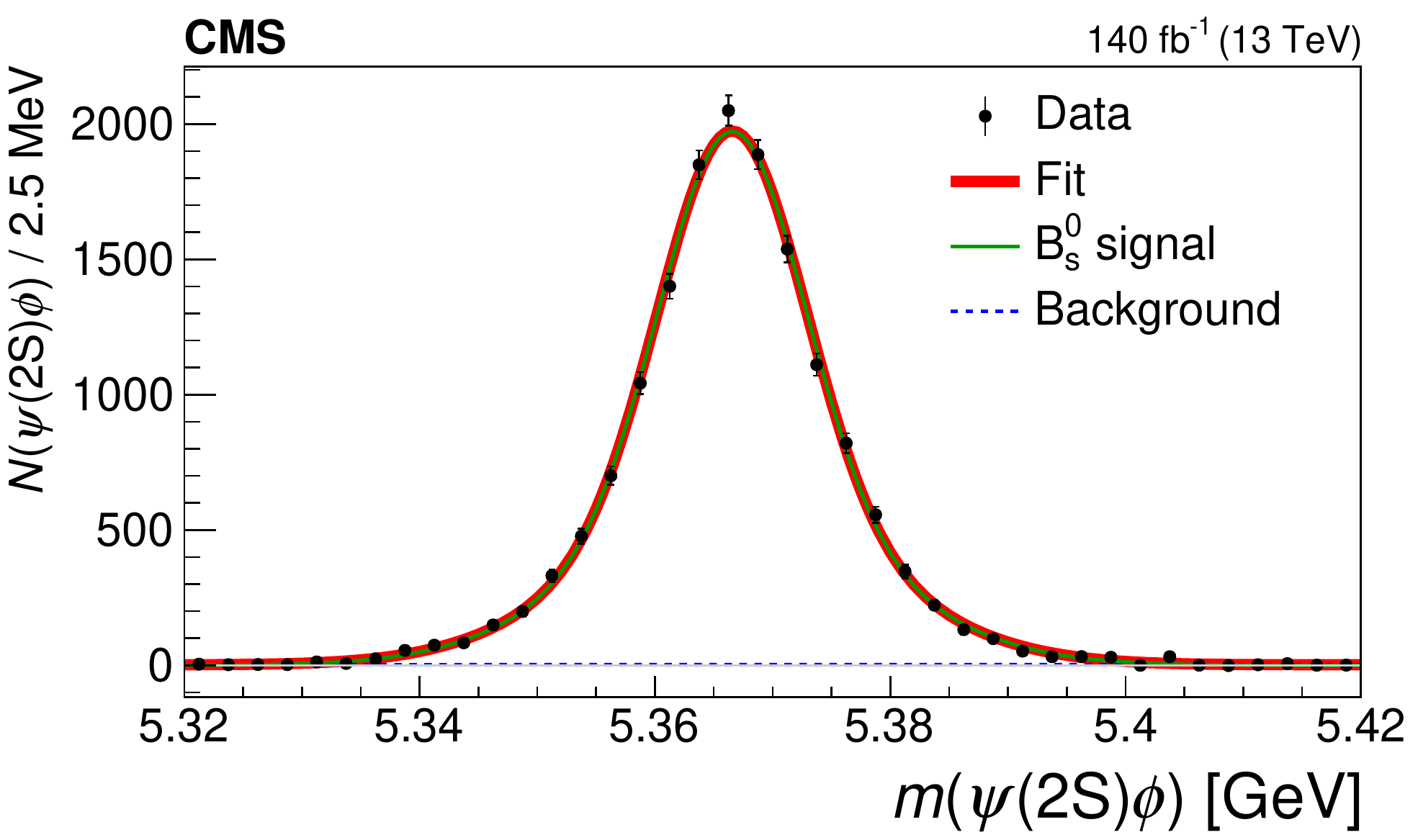}
    \includegraphics[width=0.48\textwidth]{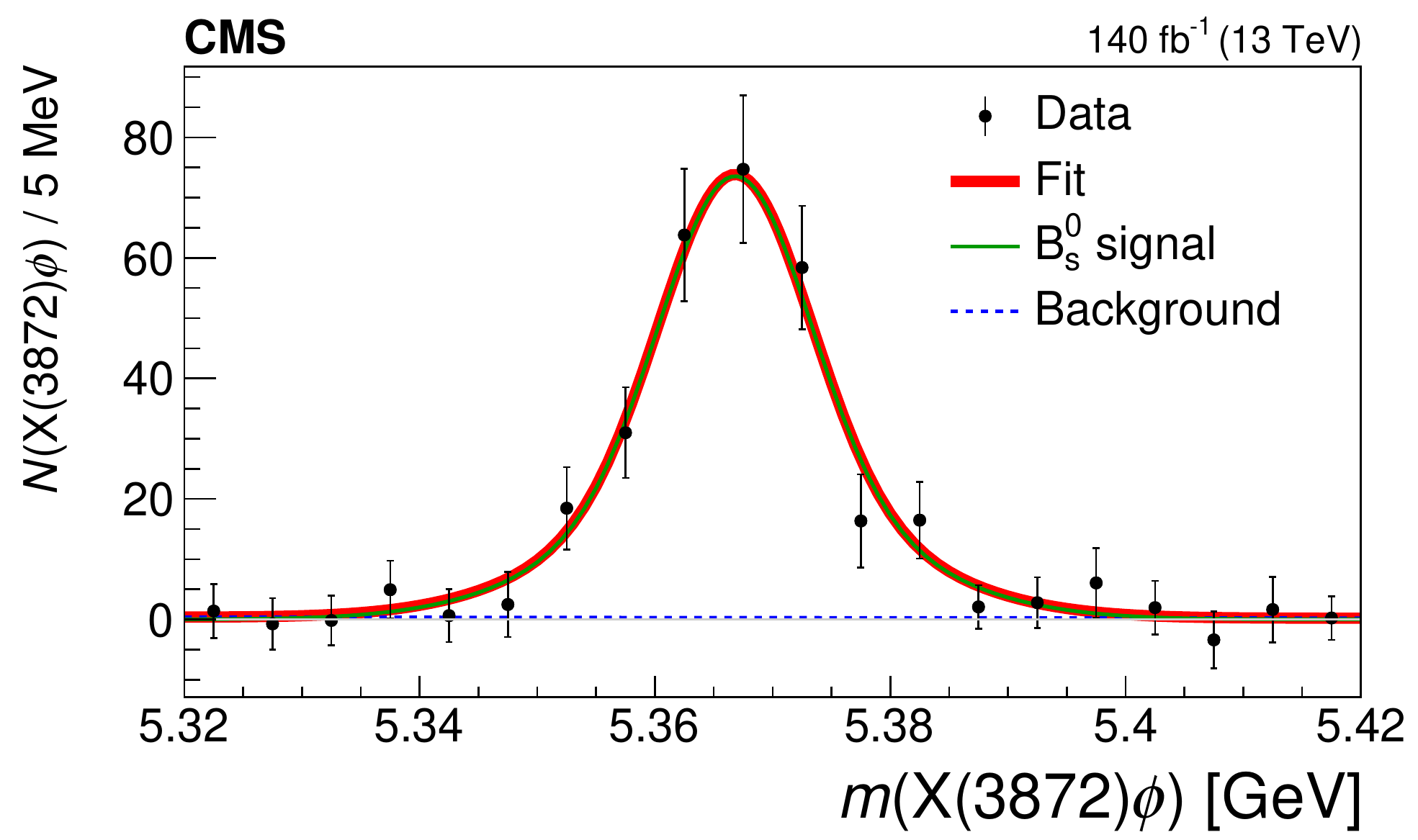}
    \caption{Background-subtracted $\psit\PGf$ (\cmsLeft) and 
    $\Xm\PGf$ (\cmsRight) invariant mass distributions obtained 
    by \splot weighting. The result 
    of each fit and its components are shown by the lines.
\label{fig:SplottedMB}}
\end{figure}

{\tolerance=800
To evaluate the background contribution related to the non-$\Bs$
production of $\psit\PGf$ in the mass range $5.32<m(\psit\PGf)<5.42\GeV$, 
the mass distribution of $\psit\PGf$ is studied, as shown in 
Fig~\ref{fig:SplottedMB} (\cmsLeft). 
The background-subtraction technique \splot~\cite{Pivk:2004ty} is used, 
together with the 2D fit described above, to subtract backgrounds 
from the nonresonant \KK and \JPP combinations.
The observed $m(\psit\PGf)$ distribution is
fitted with a DG function for the signal and an exponential for the background, as
shown in Fig.~\ref{fig:SplottedMB} (\cmsLeft).
The fit returns a non-$\Bs$ background contribution of 0.5\%. 
The same procedure is repeated in the $\Xm\PGf$ channel, shown 
in Fig.~\ref{fig:SplottedMB} (\cmsRight), and the measured contribution 
of the non-$\Bs$ background is 1.7\%.
Thus, the ratio of the event yields $\Xm/\psit$ 
changes by 1.2\% after accounting for this background from the non-$\Bs$ production 
of $\psit\PGf$ and $\Xm\PGf$ combinations.
The significance of the \BsXP signal extracted from the binned fit to the
background-subtracted $m(\Xm\PGf)$ distribution exceeds $10\sigma$.
\par}

The efficiencies for the signal and normalization channels are calculated
using the simulated event samples. The total efficiency includes 
the detector acceptance, trigger, and candidate reconstruction efficiencies.
Only the ratio of the efficiencies for the \psit and \Xm decay modes is needed 
to calculate the ratio $R$, which eliminates the systematic uncertainties related to 
the track and muon reconstruction. 
The obtained efficiency ratio is 
$\epsilon_{\BsPP}/\epsilon_{\BsXP} = 1.136\pm0.026$. It is larger than unity due to 
a tighter requirement on the dipion mass $m(\pipi)>0.7\GeV$, applied in the \Xm channel.
The reported uncertainty is related to the size of the simulated samples.
The simulated event samples are validated by comparing distributions of variables used in 
the candidate selection between the background-subtracted data and simulation. 
As no significant deviation is found, no additional systematic
uncertainty in the efficiency ratio is assigned.

Several sources of systematic uncertainty in the measured ratio $R$ are considered.
To evaluate the systematic uncertainties related to the choice of the fit model, 
several alternative functions are tested.
Uncertainties related to the choice of the signal and background models are
calculated separately. 

The systematic uncertainty in the modeling of the \PKK signal is estimated by varying the 
$\PGf$ natural width and the \mKK resolution within their uncertainties.
The corresponding changes in the ratio $R$ are negligible.
The systematic uncertainty in the \mKK and \mJPP background model is estimated 
by testing alternative models. Instead of the baseline model, 
either a second-order polynomial or a threshold function multiplied 
by this polynomial is used.
The systematic uncertainty in the \JPP signal model is estimated by replacing 
the DG function with a Student's t-distribution~\cite{Studentt} or, 
for the \Xm channel, by conservatively scaling the resolution obtained 
in the $\psit$ channel by the ratio of the resolutions of the two 
channels observed in the simulation.

The systematic uncertainty related to the non-$\Bs$ background 
is estimated using the \splot technique to subtract the contributions from 
nonresonant \KK and \JPP combinations from the $m(\Bs)$ distribution, 
as described above and shown in Fig.~\ref{fig:SplottedMB}. 
A systematic uncertainty of 1.2\% is assigned, based 
on the fit results to the background-subtracted $m(\psit\PGf)$ and $m(\Xm\PGf)$
distributions.

The uncertainty related to the simulation sample size is 2.2\%, 
as evaluated above.
Changes in the detector and trigger conditions in the course of the
2016--2018 data taking are shown to have a negligible effect on the 
measured ratio, as the signal and normalization 
processes are very similar. The ratio $R$ is found to be stable 
across different years of data taking,
therefore no related systematic uncertainty is assigned.

Table~\ref{tab:syst} summarizes the systematic uncertainties described above,
together with the total systematic uncertainty, obtained by adding the effects from 
the different sources in quadrature.

\begin{table}[htb]
\centering
\topcaption{Relative systematic uncertainties in the ratio $R$.\label{tab:syst}}
\begin{scotch}{lc}
    Source                          &   Uncertainty (\%) \\  
\hline
 \mKK signal model                  &   \textless\,0.1 \\
 \mKK background model              &   2.5 \\
 \mJPP signal model                 &   5.3 \\
 \mJPP background model             &   4.3 \\
 Non-\Bs background                 &   1.2 \\
 Simulated sample size              &   2.2 \\[\cmsTabSkip]
 Total                              &   7.7
\end{scotch}
\end{table}
Using Eq.~\eqref{eqr}, together with the measured signal yields of the 
\BsXP and \BsPP decays and the corresponding efficiency ratio, the 
product of the branching fractions, with respect to that of the \BsPP decay, is
measured to be
\begin{linenomath}\begin{equation*}
R = (2.21\pm0.29\stat\pm0.17\syst)\%.
\end{equation*}\end{linenomath}

{\tolerance=2400
Multiplying the measured ratio $R$ by the known branching fractions
$\mathcal{B}(\BsPP)$
and
$\mathcal{B}(\PJPP)$~\cite{PDG}, we obtain
\ifthenelse{\boolean{cms@external}}
{
$\mathcal{B}(\Bs\to\Xm\PGf)\,\mathcal{B}(\XmJPP) =
(4.14\pm0.54\stat\pm0.32\syst\pm0.46\ebran)\times10^{-6}$,
}
{
\begin{linenomath}\begin{equation*}
\mathcal{B}(\Bs\to\Xm\PGf)\,\mathcal{B}(\XmJPP) =
(4.14\pm0.54\stat\pm0.32\syst\pm0.46\ebran)\times10^{-6},
\end{equation*}\end{linenomath}
}
where the last uncertainty is related to the uncertainties in the 
aforementioned world-average branching fractions.
\par}

{\tolerance=4800
This branching fraction product can be compared to similar 
ones in \Bd and \Bu decays~\cite{PDG}: 
$\mathcal{B}(\Bd\to\Xm\Kzb)\mathcal{B}(\XmJPP) =
(4.3\pm1.3)\times10^{-6}$ and 
$\mathcal{B}(\Bu\to\Xm\Kp)\mathcal{B}(\XmJPP) =
(8.6\pm0.8)\times10^{-6}$.
The measured value for \Bs is consistent with that for \Bd but
about two times smaller than the one for \Bu:
\ifthenelse{\boolean{cms@external}}
{
$\mathcal{B}(\BsXP)/\mathcal{B}(\Bu\to\Xm\Kp) =
0.482\pm0.063\stat\pm0.037\syst\pm0.070\ebran$.
}
{
\begin{linenomath}\begin{equation*}
\frac{\mathcal{B}(\BsXP)}{\mathcal{B}(\Bu\to\Xm\Kp)} =
0.482\pm0.063\stat\pm0.037\syst\pm0.070\ebran.
\end{equation*}\end{linenomath}
}
This ratio is significantly lower than the corresponding 
one for decays to the charmonium state $\psit$ of
$\mathcal{B}(\BsPP)/\mathcal{B}(\Bu\to\psit\Kp)=0.87\pm0.10$~\cite{PDG}.
While this work was in the journal review, an explanation of the observed 
difference in the decay branching fractions has been proposed~\cite{Maiani:2020zhr}
within the tetraquark model of the \Xm state.
\par}

In summary, using a data sample corresponding to an integrated luminosity of 
$140\fbinv$ of proton-proton collisions collected by the CMS experiment
at $\sqrt{s}= 13\TeV$ in 2016--2018, the \BsXP decay is observed for 
the first time.
The comparison with similar decays of \Bd and \Bu mesons indicates that the \Xm 
formation in \PB meson decays is different from \psit formation, suggesting that 
\Xm is not a pure charmonium state, supporting similar conclusions derived 
from other experimental 
measurements~\cite{lhcb,Aaij:2015eva,Abulencia:2006ma,allaboutrho,Chatrchyan:2013cld,Aaboud:2016vzw,Brambilla:2010cs}.
This observation may shed new light on the nature of the \Xm particle.

\begin{acknowledgments}
We congratulate our colleagues in the CERN accelerator departments for the excellent performance of the LHC and thank the technical and administrative staffs at CERN and at other CMS institutes for their contributions to the success of the CMS effort. In addition, we gratefully acknowledge the computing centers and personnel of the Worldwide LHC Computing Grid for delivering so effectively the computing infrastructure essential to our analyses. Finally, we acknowledge the enduring support for the construction and operation of the LHC and the CMS detector provided by the following funding agencies: BMBWF and FWF (Austria); FNRS and FWO (Belgium); CNPq, CAPES, FAPERJ, FAPERGS, and FAPESP (Brazil); MES (Bulgaria); CERN; CAS, MoST, and NSFC (China); COLCIENCIAS (Colombia); MSES and CSF (Croatia); RPF (Cyprus); SENESCYT (Ecuador); MoER, ERC IUT, PUT and ERDF (Estonia); Academy of Finland, MEC, and HIP (Finland); CEA and CNRS/IN2P3 (France); BMBF, DFG, and HGF (Germany); GSRT (Greece); NKFIA (Hungary); DAE and DST (India); IPM (Iran); SFI (Ireland); INFN (Italy); MSIP and NRF (Republic of Korea); MES (Latvia); LAS (Lithuania); MOE and UM (Malaysia); BUAP, CINVESTAV, CONACYT, LNS, SEP, and UASLP-FAI (Mexico); MOS (Montenegro); MBIE (New Zealand); PAEC (Pakistan); MSHE and NSC (Poland); FCT (Portugal); JINR (Dubna); MON, RosAtom, RAS, RFBR, and NRC KI (Russia); MESTD (Serbia); SEIDI, CPAN, PCTI, and FEDER (Spain); MOSTR (Sri Lanka); Swiss Funding Agencies (Switzerland); MST (Taipei); ThEPCenter, IPST, STAR, and NSTDA (Thailand); TUBITAK and TAEK (Turkey); NASU (Ukraine); STFC (United Kingdom); DOE and NSF (USA).

\hyphenation{Rachada-pisek} Individuals have received support from the Marie-Curie program and the European Research Council and Horizon 2020 Grant, contract Nos.\ 675440, 752730, and 765710 (European Union); the Leventis Foundation; the A.P.\ Sloan Foundation; the Alexander von Humboldt Foundation; the Belgian Federal Science Policy Office; the Fonds pour la Formation \`a la Recherche dans l'Industrie et dans l'Agriculture (FRIA-Belgium); the Agentschap voor Innovatie door Wetenschap en Technologie (IWT-Belgium); the F.R.S.-FNRS and FWO (Belgium) under the ``Excellence of Science -- EOS" -- be.h project n.\ 30820817; the Beijing Municipal Science \& Technology Commission, No. Z191100007219010; the Ministry of Education, Youth and Sports (MEYS) of the Czech Republic; the Deutsche Forschungsgemeinschaft (DFG) under Germany's Excellence Strategy -- EXC 2121 ``Quantum Universe" -- 390833306; the Lend\"ulet (``Momentum") Programme and the J\'anos Bolyai Research Scholarship of the Hungarian Academy of Sciences, the New National Excellence Program \'UNKP, the NKFIA research grants 123842, 123959, 124845, 124850, 125105, 128713, 128786, and 129058 (Hungary); the Council of Science and Industrial Research, India; the HOMING PLUS program of the Foundation for Polish Science, cofinanced from European Union, Regional Development Fund, the Mobility Plus program of the Ministry of Science and Higher Education, the National Science Center (Poland), contracts Harmonia 2014/14/M/ST2/00428, Opus 2014/13/B/ST2/02543, 2014/15/B/ST2/03998, and 2015/19/B/ST2/02861, Sonata-bis 2012/07/E/ST2/01406; the National Priorities Research Program by Qatar National Research Fund; the Ministry of Science and Education, grant no. 14.W03.31.0026 (Russia); the Tomsk Polytechnic University Competitiveness Enhancement Program and ``Nauka" Project FSWW-2020-0008 (Russia); the Programa Estatal de Fomento de la Investigaci{\'o}n Cient{\'i}fica y T{\'e}cnica de Excelencia Mar\'{\i}a de Maeztu, grant MDM-2015-0509 and the Programa Severo Ochoa del Principado de Asturias; the Thalis and Aristeia program cofinanced by EU-ESF and the Greek NSRF; the Rachadapisek Sompot Fund for Postdoctoral Fellowship, Chulalongkorn University and the Chulalongkorn Academic into Its 2nd Century Project Advancement Project (Thailand); the Kavli Foundation; the Nvidia Corporation; the SuperMicro Corporation; the Welch Foundation, contract C-1845; and the Weston Havens Foundation (USA).
\end{acknowledgments}
\bibliography{auto_generated}
\ifthenelse{\boolean{cms@external}}{}{
\clearpage
\appendix
\numberwithin{table}{section}
\numberwithin{figure}{section}
\section{Projections of two-dimensional fits\label{app:suppMat}}
The results of the 2D fit on $\mKK$ and $\mJPP$ distribution 
are shown in Fig.~\ref{fig:TwoDfitPall} for the \psit channel and in
Fig.~\ref{fig:TwoDfitXall} for the \Xm channel.  
The top row shows the fit projections on the \mKK axis in 3 ranges of \mJPP: 
left sideband, signal region, right sideband,
while the bottom row shows similar projections on \mJPP axis in 3 ranges of \mKK.

\begin{figure*}[h]
\centering
    \includegraphics[width=0.325\textwidth]{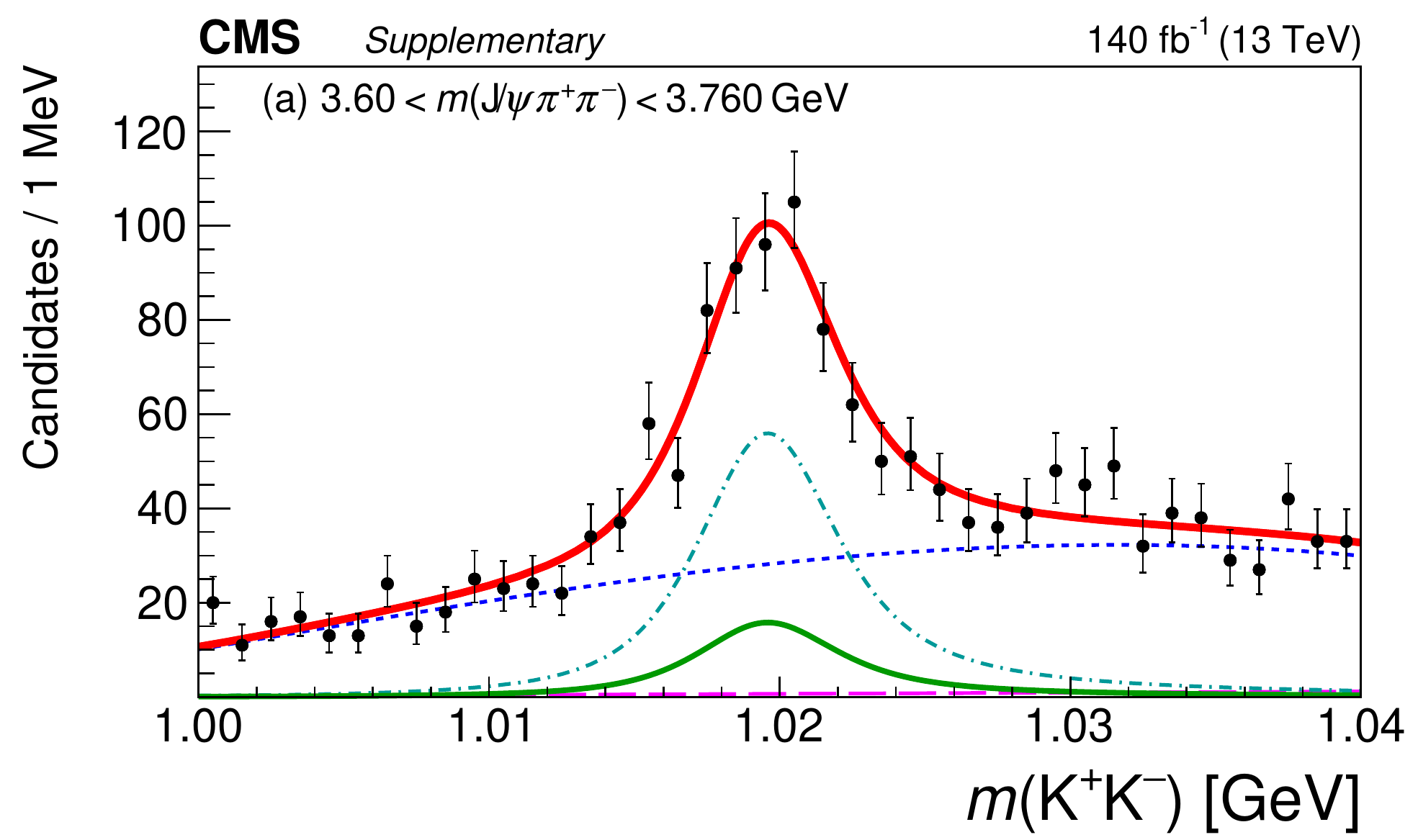}
    \includegraphics[width=0.325\textwidth]{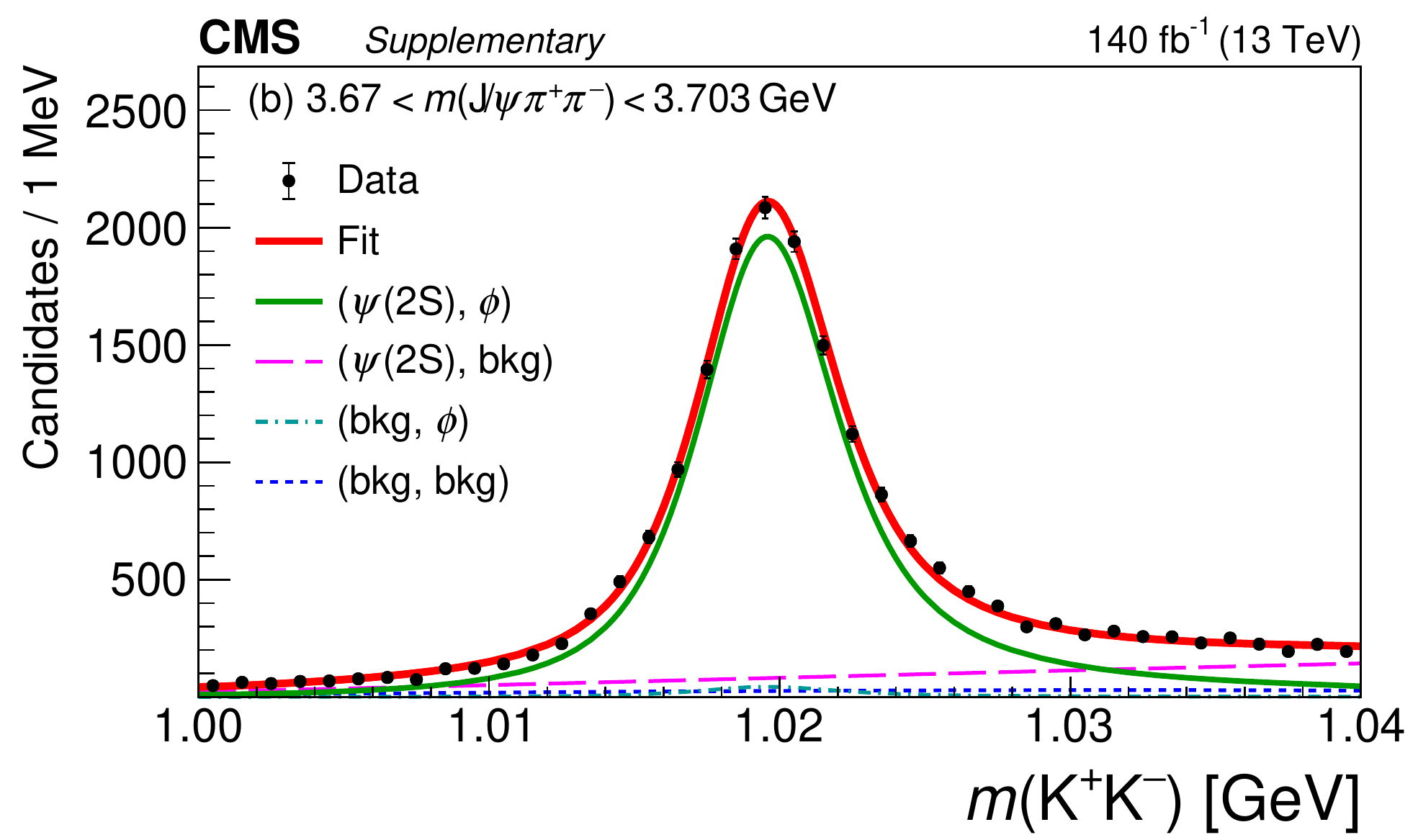}
    \includegraphics[width=0.325\textwidth]{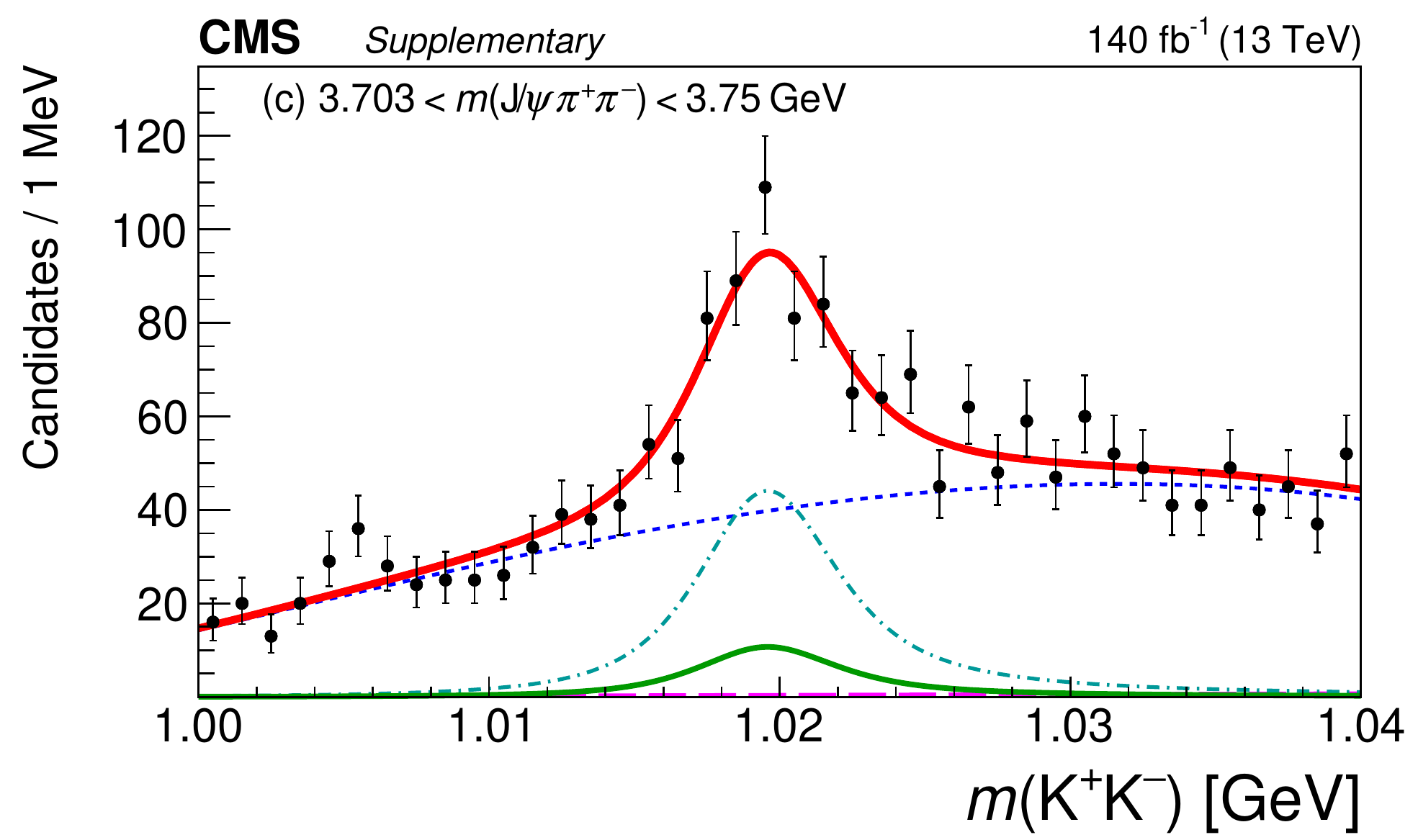}
    \includegraphics[width=0.325\textwidth]{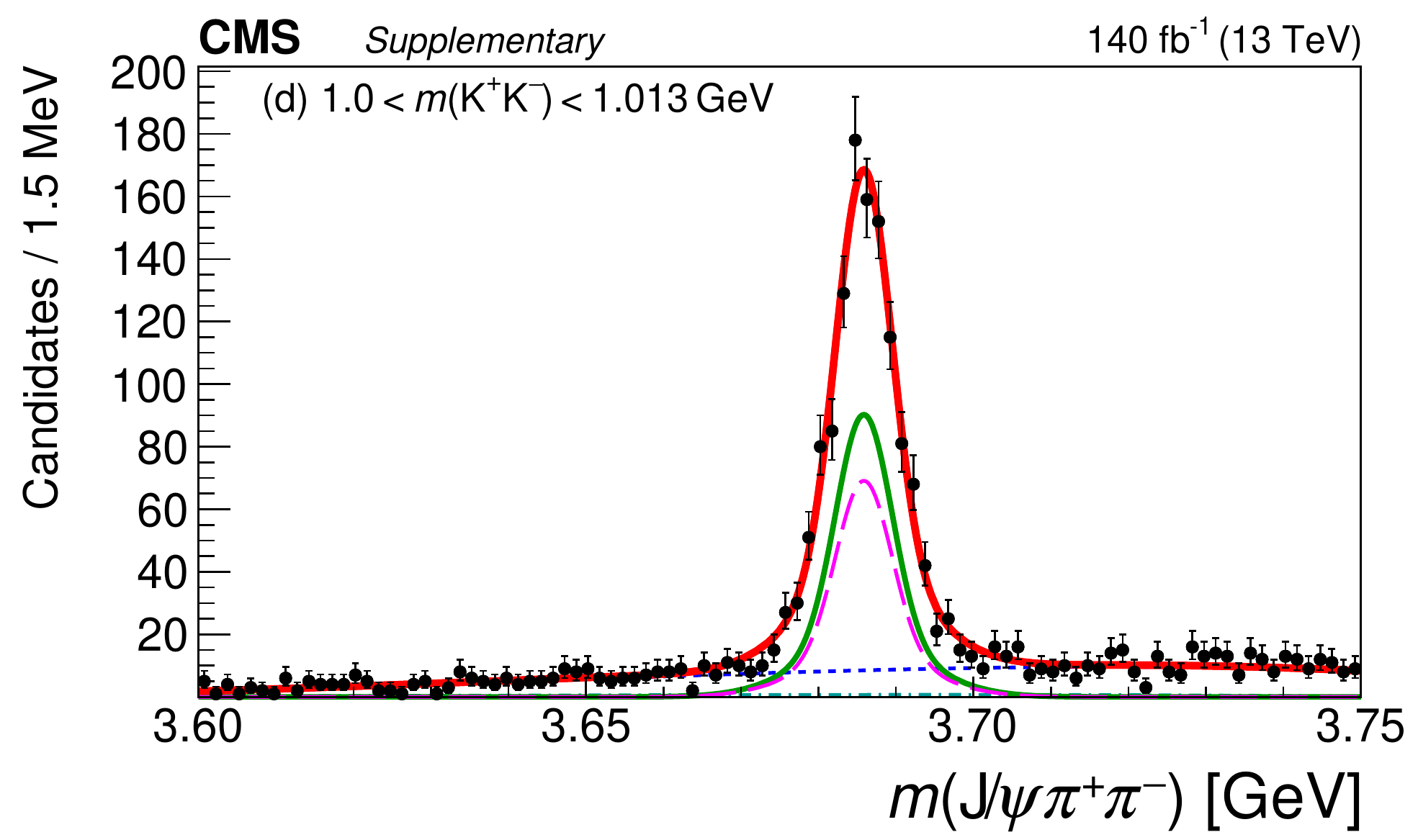}
    \includegraphics[width=0.325\textwidth]{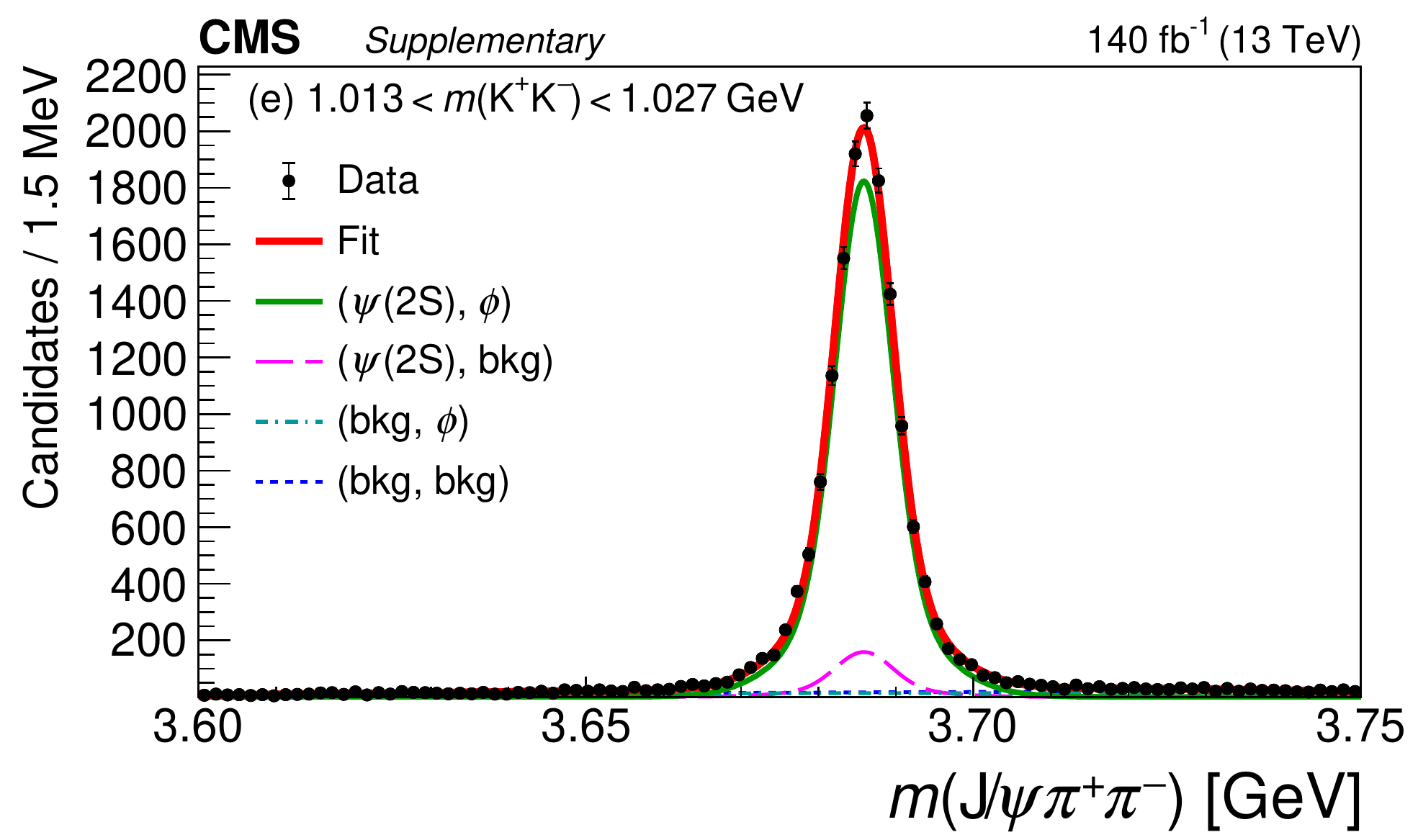}
    \includegraphics[width=0.325\textwidth]{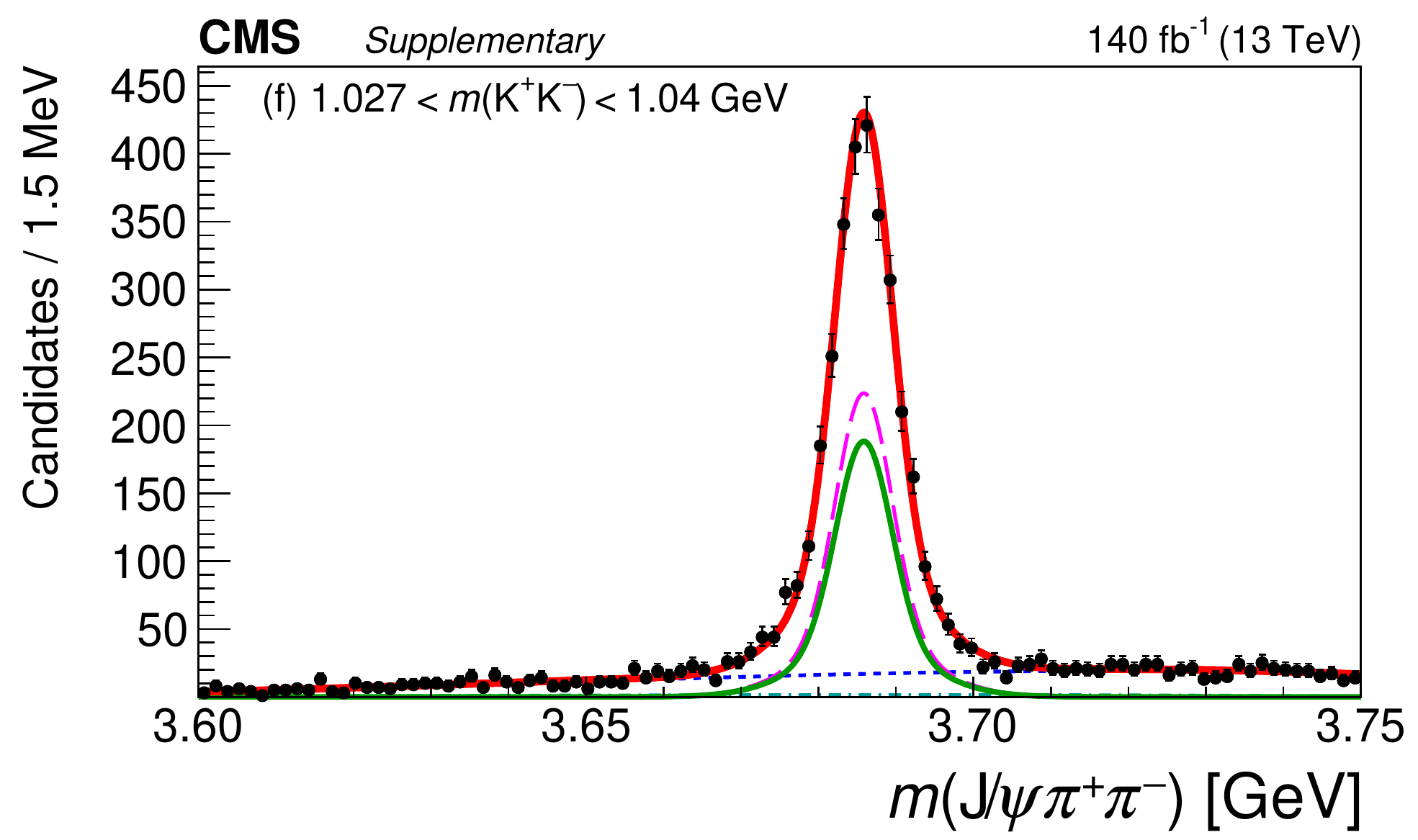}
    \caption{ Results of the 2D fit to $\mKK:\mJPP$ for \BsPP channel.
    Top row shows projections of the 2D fit on \mKK in the ranges of \mJPP: 
    left sideband (a), signal region (b), and right sideband (c). 
    Bottom row shows projections of the 2D fit on \mJPP in the ranges of \mKK: 
    left sideband (d), signal region (e), and right sideband (f). 
\label{fig:TwoDfitPall}}
\end{figure*}

\begin{figure*}[h]
\centering
    \includegraphics[width=0.325\textwidth]{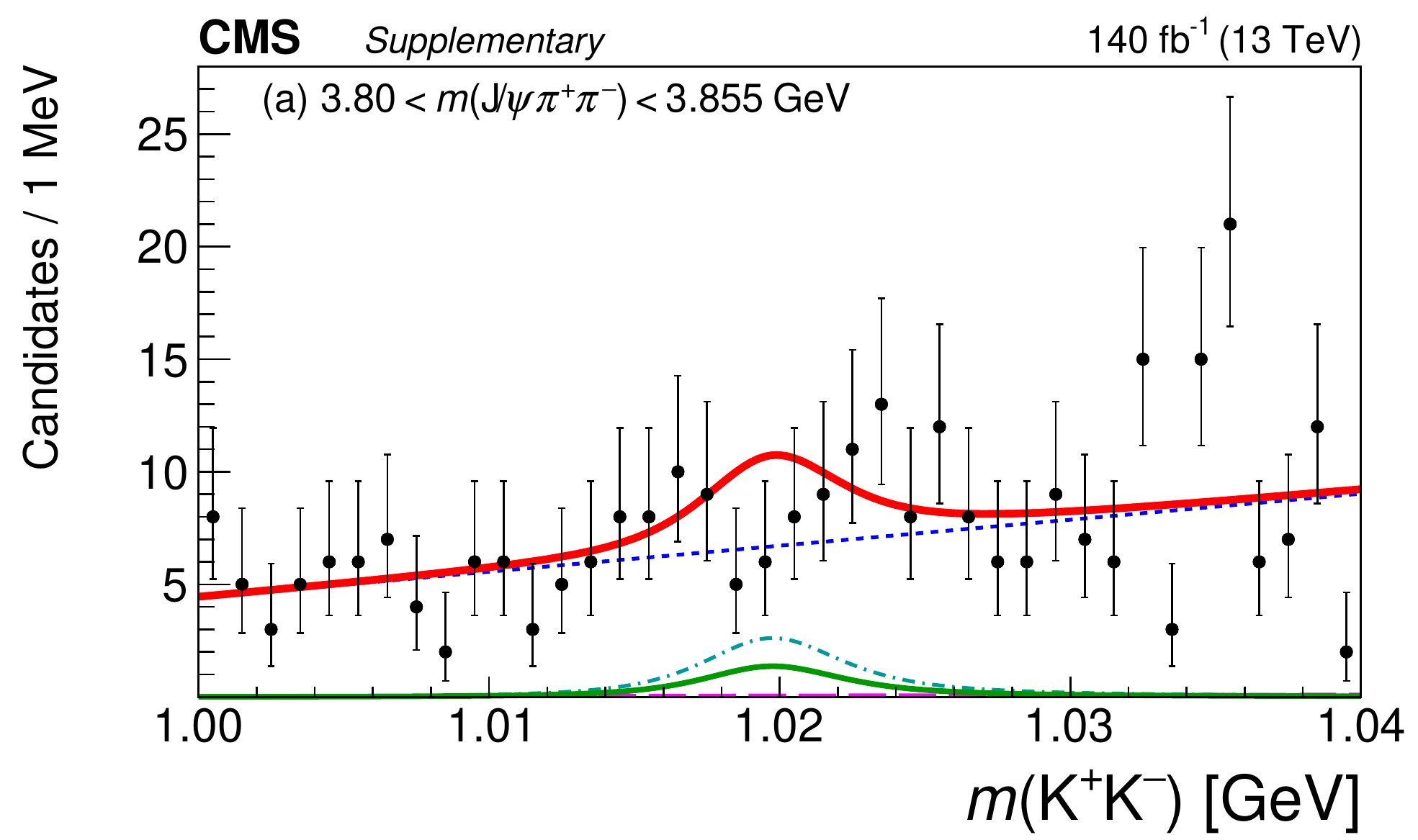}
    \includegraphics[width=0.325\textwidth]{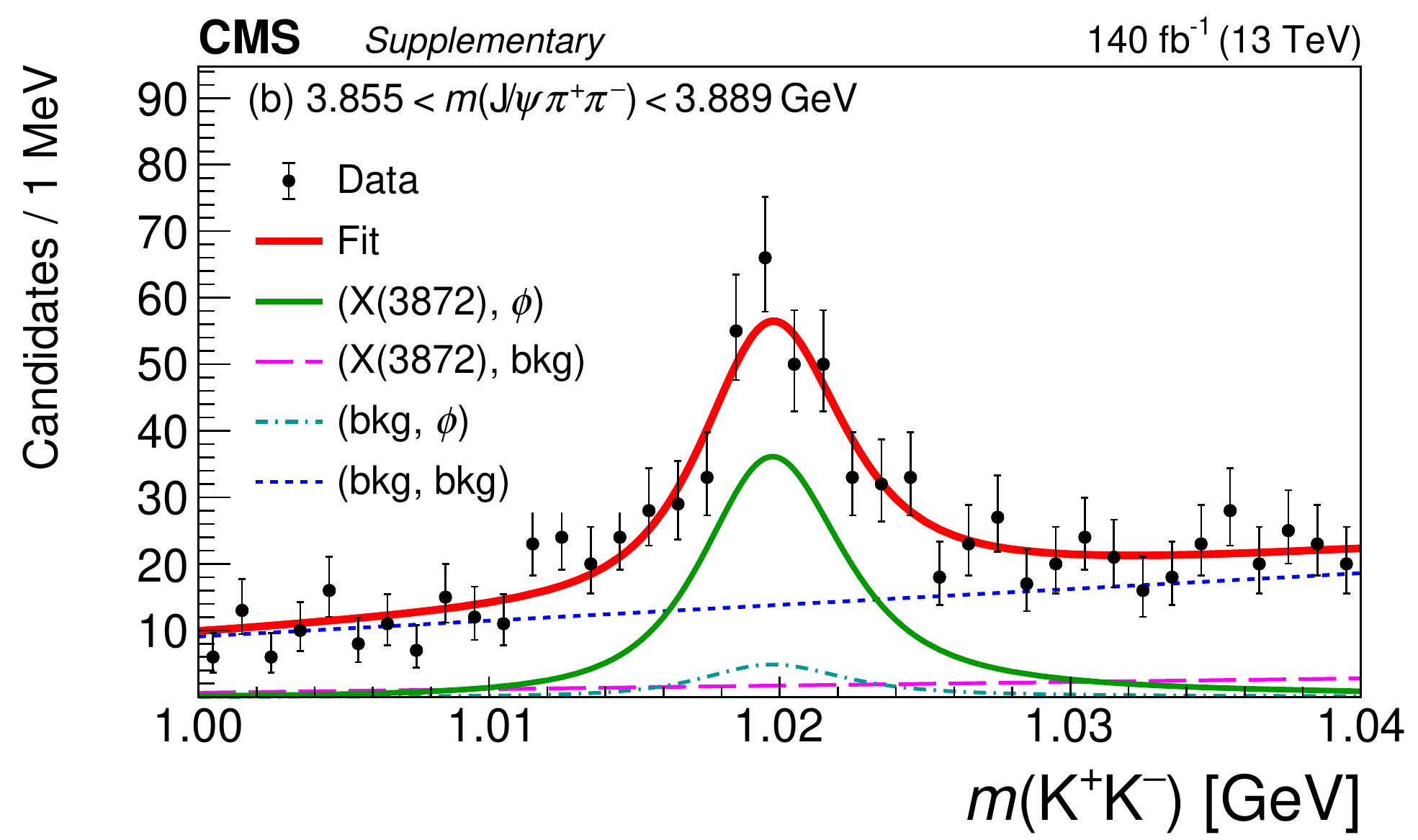}
    \includegraphics[width=0.325\textwidth]{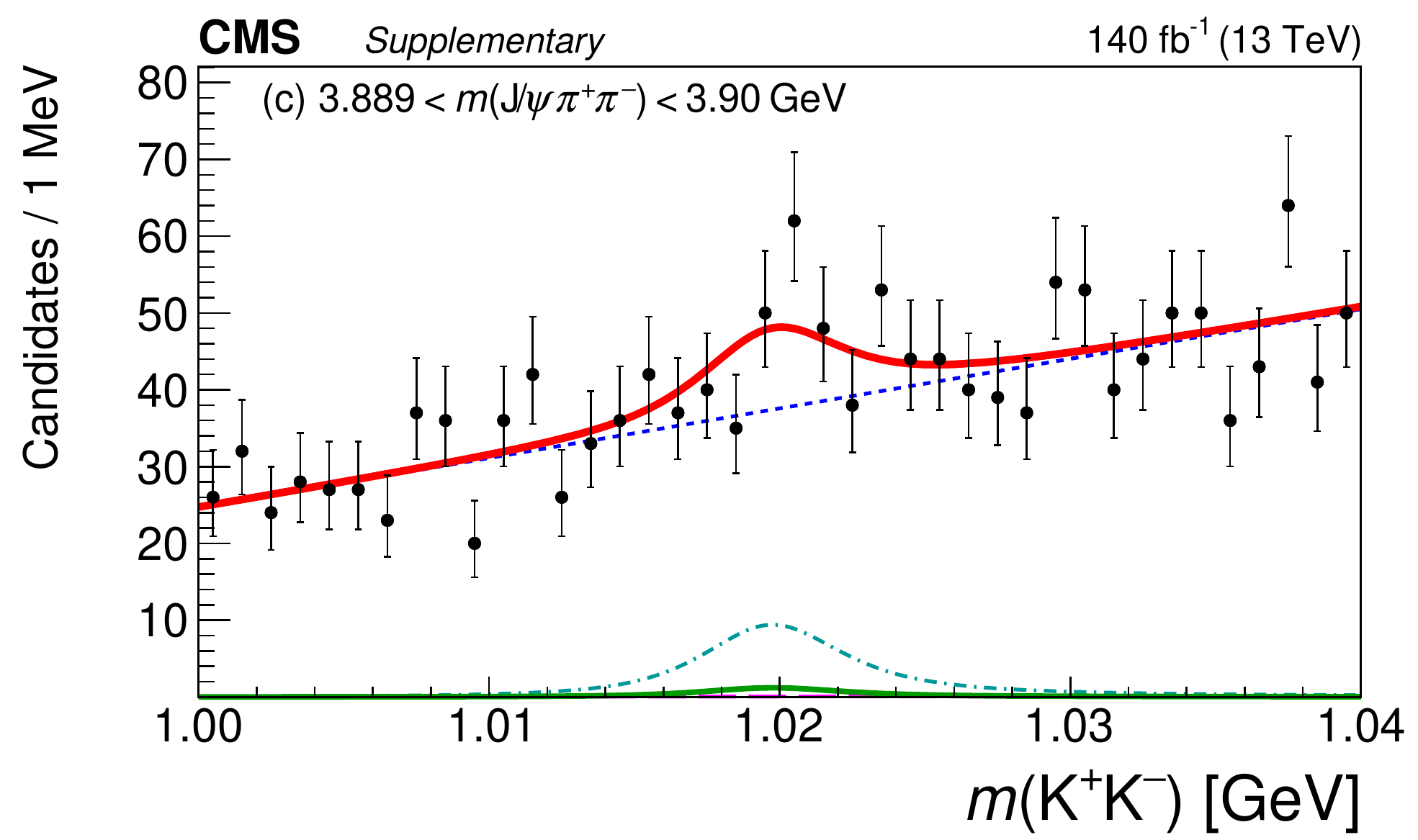}
    \includegraphics[width=0.325\textwidth]{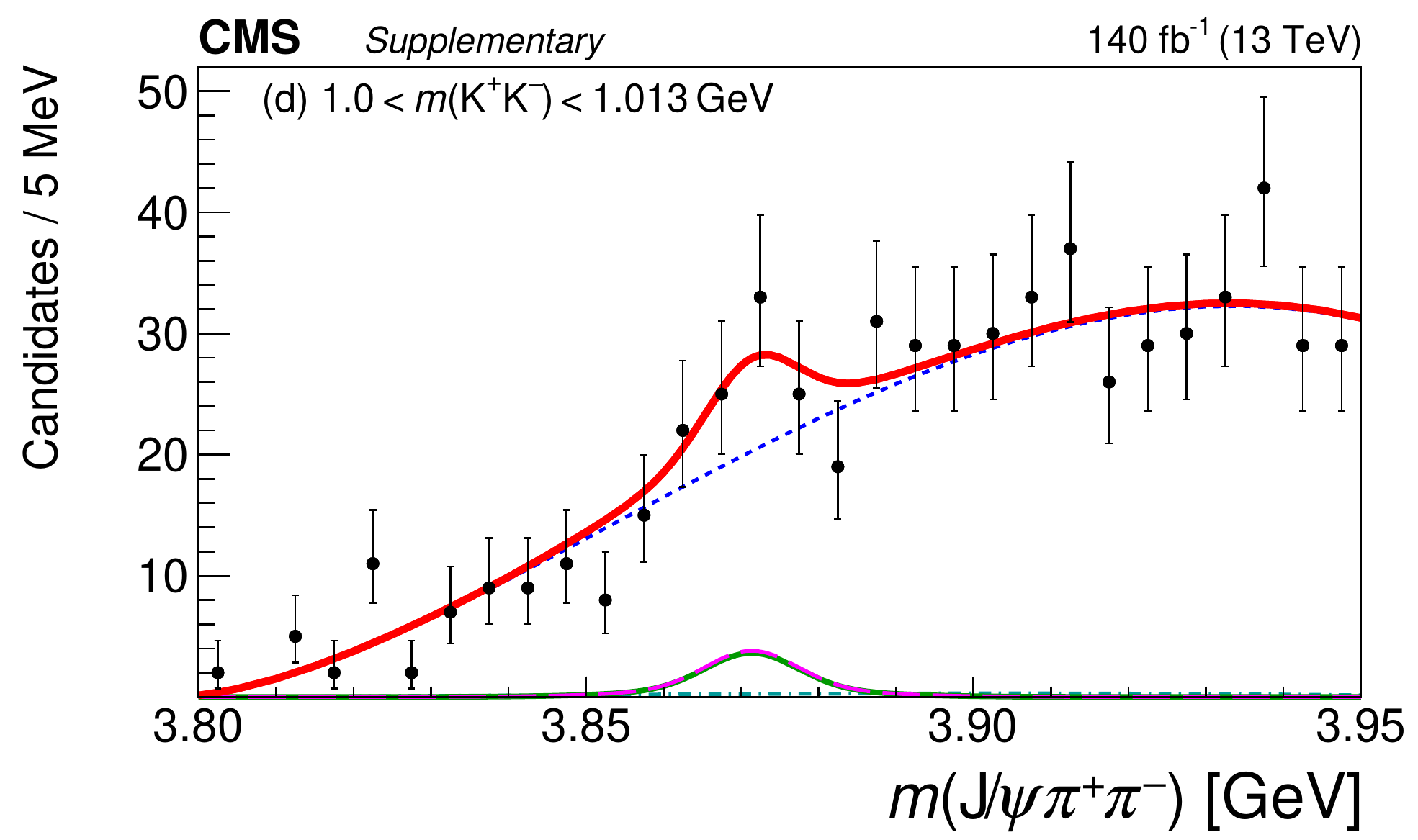}
    \includegraphics[width=0.325\textwidth]{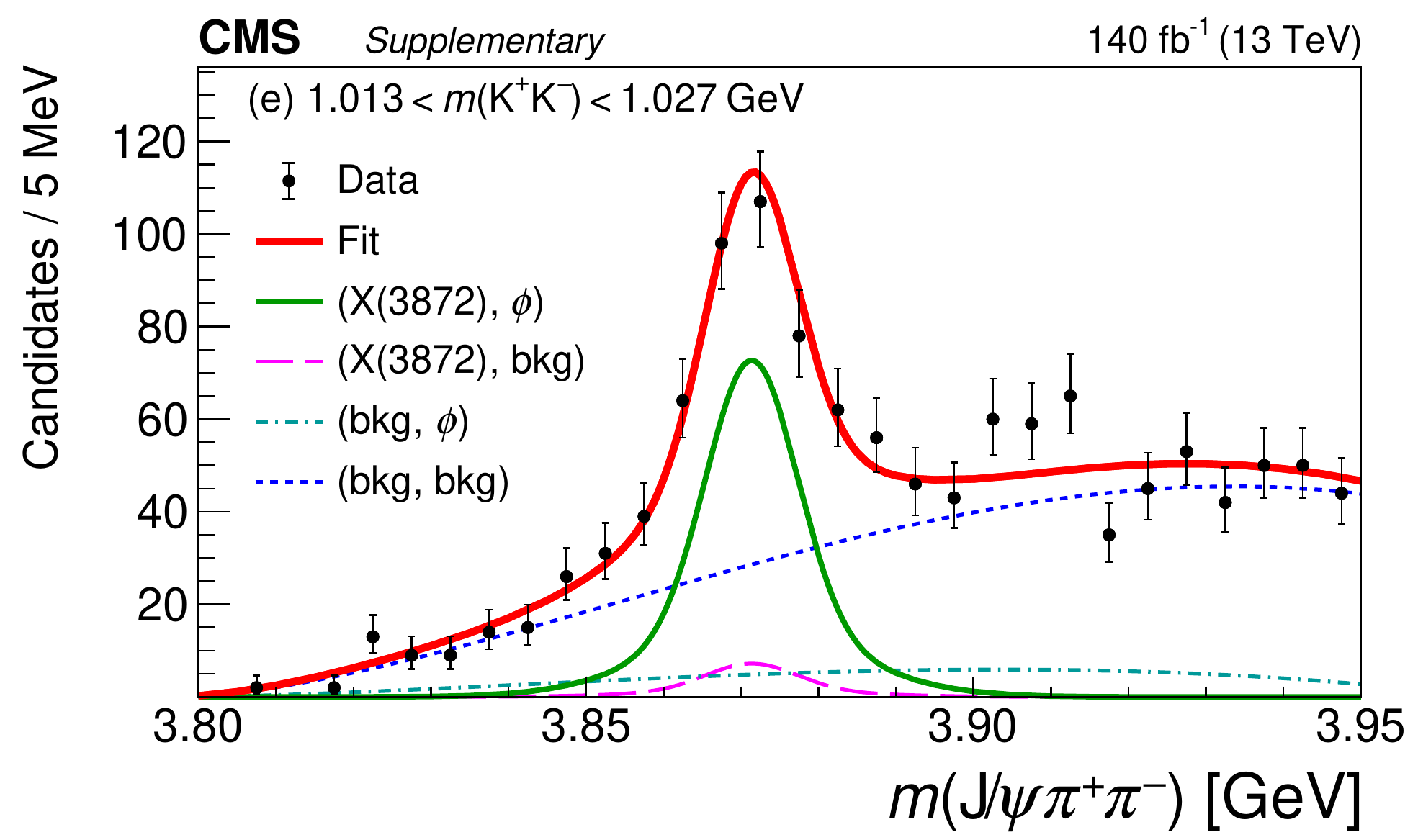}
    \includegraphics[width=0.325\textwidth]{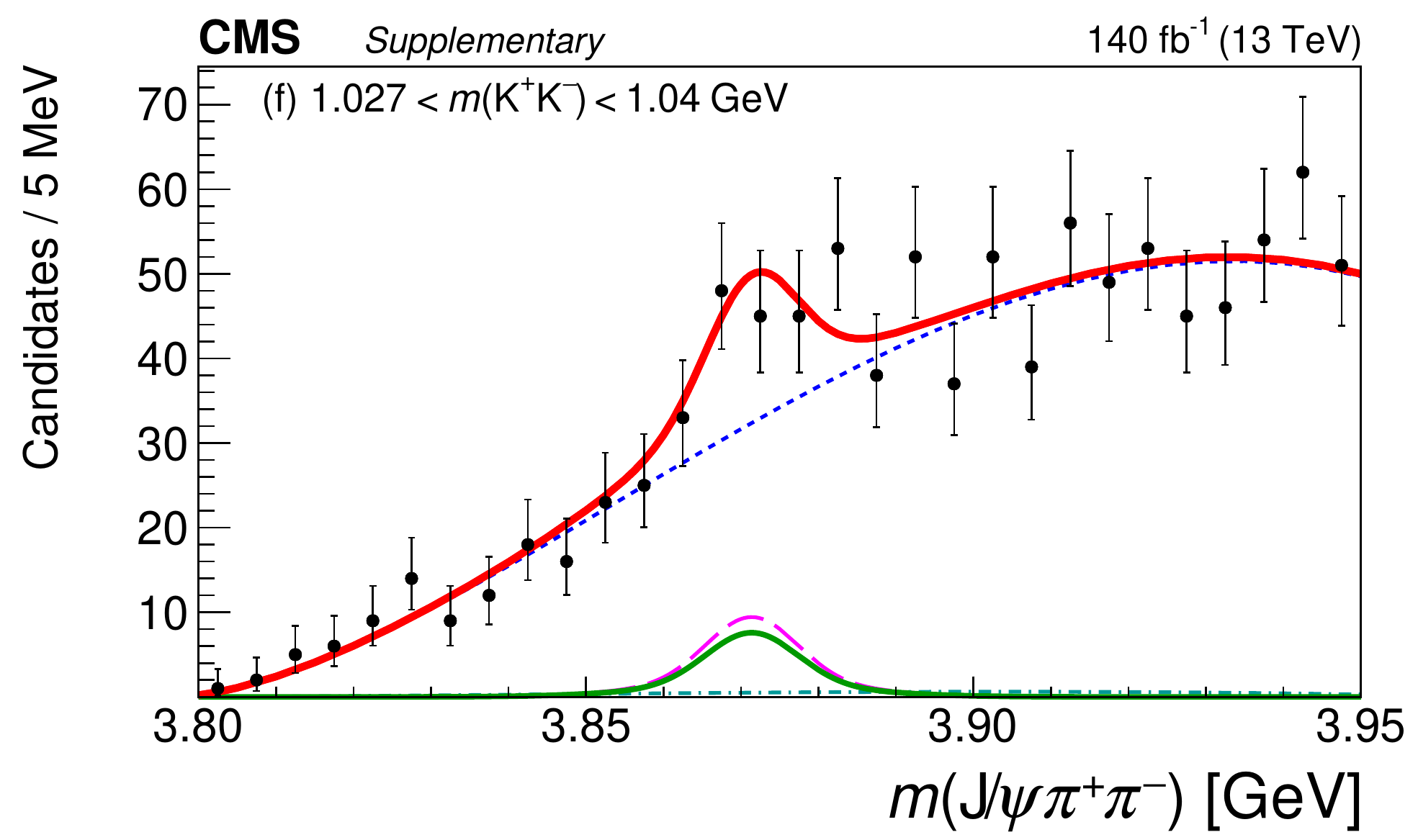}
    \caption{ Results of the 2D fit to $\mKK:\mJPP$ for \BsXP channel.
    Top row shows projections of the 2D fit on \mKK in the ranges of \mJPP: 
    left sideband (left), signal region (center), and right sideband (right). 
    Bottom row shows projections of the 2D fit on \mJPP in the ranges of \mKK: 
    left sideband (d), signal region (e), and right sideband (f). 
\label{fig:TwoDfitXall}}
\end{figure*}

}
\cleardoublepage \section{The CMS Collaboration \label{app:collab}}\begin{sloppypar}\hyphenpenalty=5000\widowpenalty=500\clubpenalty=5000\vskip\cmsinstskip
\textbf{Yerevan Physics Institute, Yerevan, Armenia}\\*[0pt]
A.M.~Sirunyan$^{\textrm{\dag}}$, A.~Tumasyan
\vskip\cmsinstskip
\textbf{Institut f\"{u}r Hochenergiephysik, Wien, Austria}\\*[0pt]
W.~Adam, F.~Ambrogi, T.~Bergauer, M.~Dragicevic, J.~Er\"{o}, A.~Escalante~Del~Valle, R.~Fr\"{u}hwirth\cmsAuthorMark{1}, M.~Jeitler\cmsAuthorMark{1}, N.~Krammer, L.~Lechner, D.~Liko, T.~Madlener, I.~Mikulec, F.M.~Pitters, N.~Rad, J.~Schieck\cmsAuthorMark{1}, R.~Sch\"{o}fbeck, M.~Spanring, S.~Templ, W.~Waltenberger, C.-E.~Wulz\cmsAuthorMark{1}, M.~Zarucki
% ...
\vskip\cmsinstskip
\dag: Deceased\\
1:  Also at Vienna University of Technology, Vienna, Austria\\
% ...
\end{sloppypar}
\end{document}